\documentclass[runningheads]{llncs}

\usepackage{caption}
\usepackage{subfigure}
\usepackage[T1]{fontenc}

\usepackage{cite}
\usepackage{amsmath,amsfonts}
\usepackage{algorithmic}
\usepackage{graphicx}
\usepackage{textcomp}
\usepackage[dvipsnames]{xcolor}
\usepackage{color, colortbl}
\definecolor{Gray}{gray}{0.9}
\usepackage{amsmath}
\usepackage{pifont}
\newcommand{\cmark}{\ding{51}}%
\newcommand{\xmark}{\ding{55}}
\usepackage{makecell}
\usepackage{wrapfig}
\usepackage{picinpar}

\usepackage[utf8]{inputenc} %
\definecolor{mydarkred}{rgb}{0.6,0,0}
\definecolor{myblue}{HTML}{268BD2}
\definecolor{mygreen}{HTML}{658354}
\definecolor{orangeinplot}{HTML}{e29c7a}
\definecolor{purpleinplot}{HTML}{7676a4}
\definecolor{greeninplot}{HTML}{288308}

\usepackage[colorlinks,
linkcolor=mydarkred,
urlcolor=RoyalBlue,
citecolor=mydarkred]{hyperref}       %
\usepackage{url}            %
\usepackage{booktabs}       %
\usepackage{amsfonts}       %
\usepackage{nicefrac}       %
\usepackage{microtype}      %

\usepackage{algorithmic}
\usepackage{graphicx}
\usepackage{textcomp}
\usepackage{xcolor}
\usepackage{multirow}
\usepackage{algorithm}
\usepackage{bm}
\usepackage{subfigure}
\usepackage{caption}

\usepackage{CJKutf8}

\usepackage[T1]{fontenc}
\usepackage{graphicx}
\usepackage[most]{tcolorbox}
\usepackage{xcolor}

\newtcolorbox[list inside=prompt]{prompt}[1][]{
    enhanced,
    colback=green!5!gray!10,
    colframe=green!20!gray!55,
    colbacktitle=green!20!gray!55,
    coltitle=white,
    fonttitle=\bfseries\sffamily,
    fontupper=\small\ttfamily,
    title=Prompt,
    attach boxed title to top center={yshift=-3pt},
    boxed title style={
        rounded corners=northeast,
        rounded corners=southeast,
        boxrule=0pt
    },
    rounded corners,
    boxrule=1.5pt,
    left=2pt,          %
    right=0pt,         %
    top=5pt,           %
    bottom=0pt,        %
    boxsep=2pt,        %
    before skip=5pt,   %
    after skip=5pt,    %
    breakable,
    #1,
}
\definecolor{greyblue}{RGB}{120,145,180}
\newtcolorbox[list inside=prompt]{encode}[1][]{
    enhanced,
    colback=greyblue!15,
    colframe=greyblue!60,
    colbacktitle=greyblue!60,
    coltitle=white,
    fonttitle=\bfseries\sffamily,
    fontupper=\small\ttfamily,
    title=Prompt,
    attach boxed title to top center={yshift=-3pt},
    boxed title style={
        rounded corners=northeast,
        rounded corners=southeast,
        boxrule=0pt
    },
    rounded corners,
    boxrule=1.5pt,
    left=2pt,          %
    right=0pt,         %
    top=5pt,           %
    bottom=0pt,        %
    boxsep=2pt,        %
    before skip=5pt,   %
    after skip=5pt,    %
    breakable,
    #1,
}

\begin{document}
\title{\textsc{ReaKase-8B}: Legal Case Retrieval via Knowledge and Reasoning Representations with LLMs}
\titlerunning{\textsc{ReaKase-8B}}
\author{Yanran Tang \and Ruihong Qiu \and Xue Li \and Zi Huang}
\authorrunning{Y. Tang et al.}
\institute{The University of Queensland\\
\email{\{yanran.tang, r.qiu, helen.huang\}@uq.edu.au, xueli@eecs.uq.edu.au}}

\maketitle              %
\begin{abstract}
Legal case retrieval (LCR) is a cornerstone of real-world legal decision making, as it enables practitioners to identify precedents for a given query case. Existing approaches mainly rely on traditional lexical models and pretrained language models to encode the texts of legal cases. Yet there are rich information in the relations among different legal entities as well as the crucial reasoning process that uncovers how legal facts and legal issues can lead to judicial decisions. Such relational reasoning process reflects the distinctive characteristics of each case that can distinguish one from another, mirroring the real-world judicial process. Naturally, incorporating such information into the precise case embedding could further enhance the accuracy of case retrieval. In this paper, a novel \textsc{ReaKase-8B} framework is proposed to leverage extracted legal facts, legal issues, legal relation triplets and legal reasoning for effective legal case retrieval. \textsc{ReaKase-8B} designs an in-context legal case representation learning paradigm with a fine-tuned large language model. Extensive experiments on two benchmark datasets from COLIEE 2022 and COLIEE 2023 demonstrate that our knowledge and reasoning augmented embeddings substantially improve retrieval performance over baseline models, highlighting the potential of integrating legal reasoning into legal case retrieval systems. The code has
been released on \href{https://github.com/yanran-tang/ReaKase-8B}{https://github.com/yanran-tang/ReaKase-8B}.

\end{abstract}

\section{Introduction}
Legal case retrieval (LCR) aims to retrieve precedents from large-scale repositories given a query case. LCR tools become an indispensable tool in modern legal practice for legal professionals such as judges and lawyers. Beyond professional practice, these tools also provide accessible resources for individuals who seek legal guidance but cannot afford costly legal services. Recent research has significantly advanced LCR methods, enabling faster and more accurate retrieval of relevant cases. These approaches are generally classified into two main paradigms: lexical retrieval models and language models (LM). Lexical models, such as BM25~\cite{BM25}, TF-IDF~\cite{TF-IDF}, and LMIR~\cite{LMIR}, focus on computing similarity scores between cases by relying on term frequency statistics. In contrast, LM–based methods~\cite{Law2Vec,Lawformer,MTFT-BERT,MVCL,BERT-PLI,LEGAL-BERT,SAILER,JOTR,DoSSIER,RPRS,NOWJ,ConversationalAgent,UA@COLIEE2022,CL4LJP} leverage pre-trained models to generate semantic representations of cases for similarity comparison. Given the complex nature of legal texts, LM-based methods have further explored similarity computation at different levels of granularity, including sentence-level~\cite{IOT-Match}, paragraph-level~\cite{BERT-PLI}, and document-level~\cite{SAILER} strategies.

Although existing methods have made progress by using raw case texts to generate high-dimensional representations for similarity computation, they often overlook two critical aspects in encoding legal cases. \textbf{(1) Legal entities relations}. Legal entities are the fundamental objects in a case, such as parties, criminal acts, and evidence. Modelling the relations among these entities provides richer structural and latent information beyond the surface text, enabling more accurate case representations. By integrating entity relations, models can produce embeddings that are semantically richer and structurally faithful to the complexity of legal texts, thereby improving the accuracy in legal case retrieval. \textbf{(2) Legal reasoning relations}. Beyond structural elements, legal cases also embody reasoning processes that link facts and issues to judicial decisions. These relations capture how courts interpret evidence, apply statutes, and weigh arguments to reach outcomes. Unlike raw text features, reasoning-based information emphasises the logical pathways that define the uniqueness of each case. Incorporating such reasoning relations into case encoding allows models to better mirror judicial decision-making, producing embeddings that more closely reflect how judges analyse cases. This will improve the distinctiveness of case representations and enhances retrieval accuracy by prioritising cases that share similar reasoning patterns with a query.

To fully exploit both legal entity relations and legal reasoning relations, this paper proposes \textsc{ReaKase-8B}, an LLM-based embedding framework that integrates knowledge and reasoning to generate more informative case representations for the legal case retrieval task. Specifically, a \textbf{legal element extraction module} is designed to identify key components of a case, including legal facts, issues, judicial decisions, and relation triplets that capture the connections between facts and issues. Further, a \textbf{legal reasoning generation module} is introduced, leveraging an LLM to produce the inferential logic linking facts and issues to judicial decisions. Building on these extracted elements, \textsc{ReaKase-8B} fine-tunes an LLM embedding model with contrastive learning, yielding a reasoning-aware case encoder. Extensive experiments on two benchmark datasets, COLIEE 2022~\cite{COLIEE2022} and COLIEE 2023~\cite{COLIEE2023}, show that \textsc{ReaKase-8B} achieves state-of-the-art performance on the LCR task. The main contributions of this paper are as follows:
\begin{itemize}
    \item \textbf{A novel knowledge and reasoning augmented embedding framework}. \textsc{ReaKase-8B} is proposed,  which is a novel LLM-based case encoder that jointly integrates legal entity relations and legal reasoning relations, addressing key limitations of prior text only embedding methods.
    \item \textbf{A novel contextualised legal case encoding}. Two modules are designed: (i) a legal element extraction module that captures legal facts, issues, decisions, and their interrelations, and (ii) a legal reasoning generation module that produces inferential logic linking facts and issues to judicial outcomes. Together, these modules enable knowledge and reasoning-aware case representations.
    \item \textbf{State-of-the-art performance on LCR}. Through contrastive fine-tuning, \textsc{ReaKase-8B} learns embeddings that align closely with legal relational knowledge and judicial reasoning. Experiments on COLIEE 2022 and COLIEE 2023 benchmarks demonstrate consistent and significant improvements over strong baselines, establishing new state-of-the-art results.
\end{itemize}

\section{Related Work}

\textbf{Legal Case Retrieval}.
In legal case retrieval, there are two main branches of methods to capture the semantic similarity between legal cases. (1) Statistical models mainly rely on using the term frequency and the inverse document frequency to measure the semantic similarity between cases. For example, TF-IDF~\cite{TF-IDF}, BM25~\cite{BM25}, and LMIR~\cite{LMIR} are typical methods in using these measurements. (2) In the language modelling era, most methods have been using advanced language models, such as BERT~\cite{BERT}, RoBERTa~\cite{RoBERTa} and MonoT5~\cite{monot5}, as legal case encoder to generate embeddings for retrieval~\cite{Law2Vec,Lawformer,MTFT-BERT,MVCL,LEGAL-BERT,JOTR,DoSSIER,RPRS,NOWJ,UA@COLIEE2022,IOT-Match,Law-Match,LEDsummary,BM25injtct,LeiBi,LEVEN,JNLP@COLIEE2019,CL4LJP,QAjudge,query_conversational_agent,ConversationalAgent,promptcase,BERT-PLI,SAILER,lawllma,casegnn,casegnn++,caselink}. For example, BERT-PLI divides a extremely long legal case by paragraphs to obtain paragraph-level embeddings and measures the case relevance by embedding interactions between cases~\cite{BERT-PLI}. SAILER develops a fact encoder to decode reasoning and decision based on BERT and encode the case with the truncated case text~\cite{SAILER}. More recently, LLMs have been applied to introduced into embedding-based legal case retrieval by feeding the case into LLMs and use the last-layer hidden state as the case embedding, such as in LawLLM~\cite{lawllma}. Different from these methods, our proposed \textsc{ReaKase-8B} identifies the key components in a case and generate the legal reasoning and the knowledge relation triplets to support effective case retrieval with LLMs.

\vspace{0.5em}
\textbf{Embedding Models Based on Large Language Models}.
Given the powerful context comprehension ability of LLMs, recent methods have been trying to utilise LLMs to obtain meaningful document embeddings for retrieval. MMTEB benchmark serves as a testbed for different retrieval tasks, including a collection of legal question answering tasks~\cite{mmteb}. Among various methods, voyage-law-2 is the best performing models yet it is a closed-source one\footnote{https://blog.voyageai.com/2024/04/15/domain-specific-embeddings-and-retrieval-legal-edition-voyage-law-2/}. E5-Mistral-7B-Instruct~\cite{e5mistral} is one of the high performing open-sourced LLMs in legal task. Recently, Qwen3-Embedding-8B~\cite{qwen3embedding} achieves state-of-the-art embedding ability among various tasks in MMTEB. Note that the legal question answering tasks in MMTEB is not the main focus of this paper.

\vspace{0.5em}
\textbf{Legal Large Language Models}.
There are recent efforts in developing legal specific LLMs for various legal tasks~\cite{lawllma,saullm,lawllmb}. For example, general purpose legal LLMs are developed using legal domain data with supervised fine-tuning as in LawLLMs~\cite{lawllma,lawllmb}, SaulLM~\cite{saullm}. There are also frameworks directly evaluating the capability of LLM embeddings in legal scenarios by integrating LLM embeddings into various existing retrieval frameworks~\cite{priorcase,llm-sum,ns-lcr,uqlegalai}. With the prevalence in retrieval-augmented generation in LLMs, legal case retrieval has been integrated as part of the retrieval process for trustworthy legal generation~\cite{ATRIE,RELexED}. The proposed \textsc{ReaKase-8B} does not aim to build another LLM that generates token-level response to user input. But rather, the main focus of this paper is to develop a powerful legal embedding model for legal case retrieval.

\begin{figure}[!t]
\centering
\includegraphics[width=\textwidth]{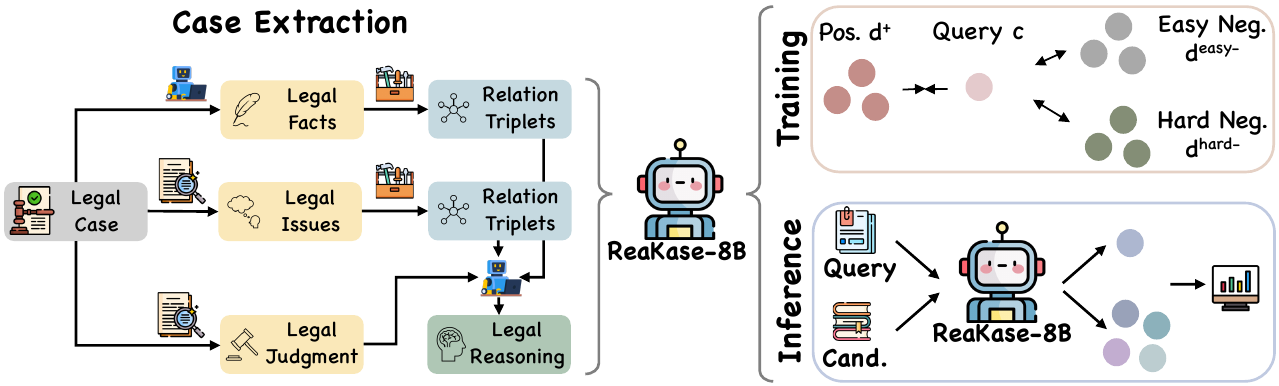}
\caption{\textsc{ReaKase-8B} designs key information extraction for knowledge triplets and legal reasoning to obtain contextualised case representations. The training is based on contrastive learning with positive (Pos.), easy negative (Neg.) and hard negative samples given a query case. During inference, \textsc{ReaKase-8B} embeddings for the query case and candidate (Cand.) cases will be used for retrieval.}
\label{fig:reakase}
\end{figure}

\section{Method}
This section presents \textsc{ReaKase-8B} in detail with the diagram in Figure~\ref{fig:reakase}.

\vspace{0.5em}
\noindent\textbf{Task Definition}.
In the legal case retrieval task, the goal is to identify precedents relevant to a given query case $q$ from a collection of candidate cases. Formally, let $\mathcal{D} = \{d_1, d_2, \dots, d_n\}$ denote a repository of $n$ candidate cases. The objective is to retrieve a subset of relevant cases, defined as  
$\mathcal{D}^* = \{ d_i \mid d_i \in \mathcal{D}, \; \text{relevant}(d_i, q) \}$, where $\text{relevant}(d_i, q)$ indicates that case $d_i$ is legally relevant to the query case $q$.

\subsection{Legal Element Extraction}
\label{sec:extract}
To capture the key legal elements that characterise a case, legal facts, issues, and judgements are extracted by leveraging legal knowledge from well-structured legal case documents.

\vspace{0.5em}
\noindent\textbf{Legal Facts.}
Legal facts capture the essential elements of a case, describing the ``who, when, what, where, and why''. For example, in legal cases from COLIEE, these details are typically located in the Background section $c_{\text(Bg)}$, which often spans thousands of words and includes redundant or repetitive content. Such noise hinders the generation of effective case representations. To mitigate this, we employ GPT-5 with the prompt in below to generate concise and accurate summaries of case facts $c_{\text{Fact}}$ with the Background:
\begin{center}
\begin{prompt}[title={Prompt Template for Legal Fact Extraction}, label=prompt:recon]
\vspace{-1mm}
\setlength{\parindent}{2em} 
\texttt{Summarize in 50 words:} $\{c_{\text(Bg)}\}$
\label{fig:fact}
\end{prompt}
\end{center}

\vspace{0.5em}
\noindent\textbf{Legal Issues.}
In the legal domain, an issue refers to ``a critical feature that focuses on the dispute points between the parties in the case''~\footnote{https://www.uscourts.gov/glossary}. In the COLIEE datasets, issues typically appear in the Analysis section, where judges articulate the contested points and provide supporting explanations. In this process, facts, issues, or judgements from precedents are often cited to justify the final decision. To anonymise these references, the datasets replace cited case names with placeholders such as \texttt{FRAGMENT\_SUPPRESSED}. These placeholders indicates that there is a reference of precedent case in this sentence based on the legal logic presented in the case text. Following this design, all sentences containing such placeholders are regarded as the legal issues $c_{\text{Issue}}$ of the case $c$, as they explicitly capture both the disputes under consideration and the precedents invoked in judicial reasoning. The extraction legal issues are then defined as:
\begin{equation}
\label{eq:issue_simple}
c_{\text{Issue}} = \{\, s \in c_{\text{Ana}} \mid \text{[PH]} \in s \,\},
\end{equation}
where $c_{\text{Ana}}$ denotes the set of sentences in the Analysis section of case $c$, and $s$ is a sentence. PH is the placeholder and it can be special tokens, such as \texttt{FRAGMENT\_SUPPRESSED}.

\vspace{0.5em}
\noindent\textbf{Legal Judgements.}
In legal cases, the judgement represents the court’s final decision, reflecting both the resolution of the dispute and the authoritative determination of the legal issues. Within the COLIEE datasets, judgements are typically located in the concluding section of case documents, often signposted by headings such as ``Judgement'' or “Order.” This consistent formatting provides a reliable basis for automatically extracting the judgement text. However, in some cases, editor names or attribution notes appear after the true ending of the judgement, which are not part of the judicial decision. To maintain accuracy and relevance, such irrelevant content is removed during preprocessing. The resulting judgement text for each case $c$ is denoted as $c_{\text{Jud}}$:
\begin{equation}
\label{eq:judgement}
c_{\text{Jud}} = \{\, s \in c_{\text{Con}} \mid s \ \text{follows headings ``Judgement'' or ``Order''} \,\},
\end{equation}
where $c_{\text{Con}}$ denotes the set of sentences in the concluding section of the case $c$.

\subsection{Legal Relation Triplets}
To capture structural information in legal texts, we extract relational triplets from the previously identified legal facts and issues. This extraction leverages open-source named entity recognition (NER) and relation extraction tools, which identify entities in the text and the semantic relations between them. For instance, in the COLIEE datasets derived from the Federal Court of Canada~\cite{COLIEE2022,COLIEE2023}, the sentence ``The claimant filed an appeal'' yields the triplet $(\textit{claimant},\ \textit{filed},\ \textit{an appeal})$. Formally, a case is represented as a set of relational triplets $\mathcal{R} = {(h, r, t)_{i=1:n}}$, where $h$ and $t$ denote the head and tail entities, $r$ denotes the relation, and $n$ is the total number of extracted triplets. Specifically, we denote the triplets extracted from legal facts and legal issues as $\mathcal{R}_{\text{Fact}}$ and $\mathcal{R}_{\text{Issue}}$, respectively. Modelling the relations among diverse legal entities captures richer structural and latent information than raw text alone, providing complementary semantic and structural signals to enhance case representations for downstream tasks.

\subsection{Legal Reasoning Generation}
Legal reasoning refers to the logical process through which a final judgement is derived from the underlying legal facts and issues of a case. To generate reasoning that reflects the judicial process of judges, the previously extracted legal facts, legal issues, and legal judgements are utilised as inputs for reasoning generation. A structured prompt template is designed for this purpose:
\begin{center}
\begin{prompt}[title={Prompt Template for Generating Legal Reasoning}, label=prompt:recon]
\vspace{-1mm}
\hspace*{2em}\ \texttt{\# System Prompt}\\
\hspace*{2em}\ \texttt{Assuming you are a legal expert from Federal Court of Canada.}\\ \\
\hspace*{2em}\ \texttt{\# User Prompt}\\
\hspace*{2em}\ \texttt{Given a case with its legal facts:} $\{c_{\text{Fact}}\}$. \\
\hspace*{2em}\ \texttt{Legal fact relation triplets:} $\{R_{\text{Fact}}\}$. \\
\hspace*{2em}\ \texttt{Legal issues:} $\{c_{\text{Issue}}\}$. \\
\hspace*{2em}\ \texttt{Legal issue relation triplets:} $\{R_{\text{Issue}}\}$. \\
\hspace*{2em}\ \texttt{Final case judgement:} $\{c_{\text{Jud}}\}$. \\
\hspace*{2em}\ \texttt{Please explain how to deduce the final judgement from both legal \\ \hspace*{2em}\ facts and legal issues in 100 words.}
\label{fig:reasoning}
\end{prompt}
\end{center}
Using this prompt, the legal reasoning for each case is represented as:
\begin{equation}
\label{eq:reasoning_eq}
c_{\text{Reason}} = \text{LLM}\Big(c_{\text{Fact}};\ R_{\text{Fact}};\ c_{\text{Issue}};\ R_{\text{Issue}};\ c_{\text{Jud}}\Big),
\end{equation}
where $c_{\text{Reason}}$ denotes the generated reasoning from legal facts and issues towards the judgement of a case. ``$;$'' denotes the concatenation. GPT-5 is employed to produce concise explanations that outline the inferential steps linking legal facts and issues to the final judgement. This generated reasoning serves as a natural language interpretation of judicial decision-making, bridging the gap between raw case descriptions and outcomes. Incorporating such reasoning enhances interpretability and supports the legal case retrieval.

\subsection{\textsc{ReaKase-8B} Framework}
\label{sec:reakase}
\textsc{ReaKase-8B} is built upon a base and general embedding model, Qwen3-Embedding-8B~\cite{qwen3embedding}, which does not have natural language generation ability. The model first employs a contextualised encoding module to reconstruct each case into a structured textual input that integrates facts, issues, judgements, relation triplets, and reasoning. Building on these enriched representations, a contrastive learning objective is then utilised to fine-tune \textsc{ReaKase-8B}, ensuring the embeddings of relevant cases converge while irrelevant ones diverge. This fine-tuning process allows \textsc{ReaKase-8B} to effectively capture fine-grained semantic similarities and achieve strong generalisation across diverse legal scenarios, yielding robust and accurate precedent retrieval.

\vspace{0.5em}
\noindent\textbf{Contextualised Case Encoding}.
During training, each case is reformulated as $c_\text{Cont}$ by concatenating its extracted legal facts, legal issues, as well as their knowledge-enhanced triplets, and the generated legal reasoning:
\begin{center}
\begin{encode}[title={Prompt Template for Contextualised Case Encoding}, label=prompt:recon]
\vspace{-1mm}
\hspace*{2em}\ \texttt{\# System Prompt}\\
\hspace*{2em}\ \texttt{The following contains key components of a legal case.} \\ \\
\hspace*{2em}\ \texttt{\# User Prompt}\\
\hspace*{2em}\ \texttt{Legal facts:} $\{c_{\text{Fact}}\}$. \\ 
\hspace*{2em}\ \texttt{Legal fact relation triplets:} $\{\mathcal{R}_{\text{Fact}}\}$. \\ 
\hspace*{2em}\ \texttt{Legal issues:} $\{c_{\text{Issue}}\}$. \\ 
\hspace*{2em}\ \texttt{Legal issue relation triplets:} $\{\mathcal{R}_{\text{Issue}}\}$. \\ 
\hspace*{2em}\ \texttt{Legal reasoning:} $\{c_{\text{Reason}}\}$.
\end{encode}
\end{center}
The judgement itself is not included in this prompt template is because that the judgement itself, a fairly short decisive sentence, does not provide too much information in helping the legal case retrieval. This is also observed in empirical study that with the judgement, the retrieval performance is not improved.

Feeding the reformulated case $c_\text{Cont}$ into the LLM, the case is encoded into a dense vector representation by \textsc{ReaKase-8B} as:
\begin{equation}
\label{eq:case encoding}
    \mathbf{x}_{c}=\text{\textsc{ReaKase-8B}}(c_\text{Cont}),
\end{equation}
where $\mathbf{x}_{c}\in\mathbb{R}^d$ denotes the $d$-dimensional semantic embedding of the case, capturing its integrated factual, legal, and reasoning elements for downstream retrieval and comparison tasks.

\vspace{0.5em}
\noindent\textbf{Contrastive Learning Objective}.
The contrastive objective encourages embeddings of the query case and positive cases to be pulled closer together while pushing embeddings of negative cases apart, thereby enabling the model to learn discriminative and semantically enriched case representations. To further enhance training effectiveness, hard negative samples are introduced by retrieving cases with high BM25 relevance scores that are not labelled as ground-truth matches, ensuring the model is challenged with difficult but informative distinctions:
\begin{align}
\label{eq:cl}
  \ell = -\text{log}\frac{e^{(s(\mathbf{x}_q,\mathbf{x}_{d^+}))/\tau}}{e^{(s(\mathbf{x}_q,\mathbf{x}_{d^+}))/\tau}+\sum\limits_{i=1}^{n} e^{(s(\mathbf{x}_q,\mathbf{x}_{d^{easy-}_i}))/\tau}+\sum\limits^m_{j=1}e^{(s(\mathbf{x}_q,\mathbf{x}_{d^{hard-}_j}))/\tau}},
\end{align}
In Equation~\ref{eq:cl}, $q$ denotes the query case. The positive sample $d^+$ is derived from the ground-truth labels. Negative samples are composed of two types: easy negatives $d^{easy-}$, which are randomly drawn from the candidate pool and include in-batch samples; and hard negatives $d^{hard-}$, which are retrieved using high BM25 similarity scores. The variables $n$ and $m$ denote the number of easy and hard negative samples, respectively. The similarity function $s$, implemented using either the dot product or cosine similarity, measures the similarities between case representations. Finally, the temperature coefficient $\tau$ scales the similarity scores, controlling the sharpness of the softmax distribution during contrastive learning.

\section{Experiments}
The following research questions (RQs) are studied in this section:
\begin{itemize}
    \item \textbf{RQ1:} How does \textsc{ReaKase-8B} perform compared with existing state-of-the-art legal case retrieval models?
    \item \textbf{RQ2:} What is the impact of the extracted legal entities and the generated legal reasoning in legal case retrieval?
    \item \textbf{RQ3:} How does different prompt-based contextualisation impact legal case representations?
    \item \textbf{RQ4:} When does the generated legal reasoning help with effective legal case retrieval?
\end{itemize}

\begin{wraptable}{r}{0.5\linewidth}
\centering
\vspace{-0.8cm}
\caption{Statistics of datasets.}
\label{tab:dataset}
\resizebox{1\linewidth}{!}{
\begin{tabular}{c|cc|cc}
    \toprule
    \multirow{2}{*}{Datasets} &\multicolumn{2}{c|}{COLIEE2022} &\multicolumn{2}{c}{COLIEE2023} \\
    \cmidrule{2-5}
    &train &test &train &test \\\midrule
    \# Query &898 &300 &959 &319 \\
    \# Candidates &4415 &1563 &4400 &1335 \\
    \# Avg. relevant cases &4.68 &4.21 &4.68 &2.69 \\
    Avg. length (\# token) &6724 &6785 &6532 &5566 \\
    Largest length (\# token) &127934 &85136 &127934 &61965 \\
    \bottomrule
\end{tabular}}
\vspace{-2.3em}
\end{wraptable}
\noindent\textbf{Datasets.}
To evaluate the effectiveness of the proposed \textsc{ReaKase-8B} model, two widely adopted benchmark datasets are utilised: COLIEE2022~\cite{COLIEE2022} and COLIEE2023~\cite{COLIEE2023}. Both datasets are released as part of the Competition on Legal Information Extraction and Entailment (COLIEE) and contain legal cases from the Federal Court of Canada with statistics in Table~\ref{tab:dataset}. These two datasets constitute the most widely used English benchmarks for legal case retrieval. Beyond English, \textsc{ReaKase-8B} can be readily extended to multilingual legal systems by integrating domain-specific information extraction pipelines and multilingual language models.

\vspace{0.5em}
\noindent\textbf{Baselines.}
\label{baselines}
Three categories of baselines are included: (1) \textbf{Statistical model}: BM25~\cite{BM25}, using term frequency and inverse document frequency. (2) \textbf{Legal-specific language model (LM)}: LEGAL-BERT~\cite{LEGAL-BERT}, fine-tuning BERT with legal data; MonoT5~\cite{monot5}, being trained on broad retrieval tasks; SAILER~\cite{SAILER}, leveraging legal case structure to fine-tune BERT; PromptCase~\cite{promptcase}, reformulating the legal facts and issues in BERT. (3) \textbf{Top-performing open-sourced large language models (LLM)}: E5-Mistral-7B-Instruct (Mistral-7B)~\cite{e5mistral}, Qwen3-Embedding-8B (Qwen3-8B)~\cite{qwen3embedding} and Inf-Retriever-V1 (Inf-7B)~\cite{inf-retriever-v1} are three representative LLM embedding models on MMTEB legal tasks.

\vspace{0.5em}
\noindent\textbf{Metrics.}
\label{metrics}
We evaluate ReaKase using seven standard metrics widely adopted in information retrieval and legal case retrieval, with a focus on the top-5 retrieved cases. Following prior LCR studies~\cite{promptcase, LeCaRD, SAILER}, the metrics include Precision, Micro-F1, Macro-F1, Mean Reciprocal Rank (MRR@K), Mean Average Precision (MAP@K), and Normalized Discounted Cumulative Gain (NDCG@K). In all cases, higher values indicate better performance.

\vspace{0.5em}
\noindent\textbf{Implementation.}
\label{implementation}
For dataset preprocessing, all French text is removed from both datasets. Relation and entity extraction are conducted using open-source tools, including spaCy\footnote{\url{https://spacy.io/}}
, Stanford OpenIE~\cite{OpenIE}, and LexNLP\footnote{\url{https://github.com/LexPredict/lexpredict-lexnlp}}. Qwen3-Embedding-8B~\cite{qwen3embedding} is adopted as the base model. For training, the batch size is set to 2. We use Adam~\cite{Adam} as the optimiser, with the learning rate selected from {1e-5, 5e-6, 1e-6, 5e-7, 1e-8} and weight decay from {1e-5, 1e-4, 1e-3}. For each query, one positive sample, one randomly selected easy negative sample, and one hard negative sample are included. In-batch samples from other queries are additionally treated as easy negatives. The model input length is limited to 2048 tokens. To enable efficient fine-tuning, we employ LoRA~\cite{lora} with rank $r=8$, scaling factor $\alpha=32$, and dropout rate $0.1$ for regularising the adaptation matrices. All training experiments are run for 1,000 steps on 4 AMD Mi300x GPUs.

\subsection{Overall Performance (RQ1)}
\label{overall}
We evaluate \textsc{ReaKase-8B} against multiple baselines on COLIEE2022 and COLIEE2023 (Table~\ref{tab:overall}). \textsc{ReaKase-8B} consistently outperforms lexical methods, LM-based models, and LLM-based embeddings across nearly all metrics, demonstrating the effectiveness of legal relation entities and contrastive learning for legal case retrieval. Among traditional LMs, BM25 and PromptCase achieve moderate results, while recent LLM-based retrievers (Mistral-7B, Qwen3-8B, Inf-7B) enhance retrieval performance, attaining improved scores across all metrics. Notably, \textbf{\textsc{ReaKase-8B} generalises strongly across both datasets, outperforming all baselines} in all seven metrics, highlighting its robustness and competitive advantage in legal case retrieval.

\begin{table}[!h]\centering
\vspace{-1cm}
\caption{Results on COLIEE2022 and COLIEE2023 (\%, avg. of five runs). Boldface indicates the best method (paired t-test, $p \leq 0.05$, Bonferroni corrected).}\label{tab:overall}
\resizebox{\linewidth}{!}{
\begin{tabular}{c|l|ccccccc|ccccccc}
\toprule
\midrule
&\multirow{2}{*}{Methods} &\multicolumn{7}{c}{COLIEE2022} &\multicolumn{7}{c}{COLIEE2023}\\
\cmidrule{3-16}
&&P@5 &R@5 &Mi-F1 &Ma-F1 &MRR@5 &MAP &NDCG@5 &P@5 &R@5 &Mi-F1 &Ma-F1 &MRR@5 &MAP &NDCG@5\\\midrule
&BM25 &17.9 &21.2 &19.4 &21.4 &23.6 &25.4 &33.6 &16.5 &30.6 &21.4 &22.2 &23.1 &20.4 &23.7\\
\midrule
\multirow{4}{*}{\rotatebox{90}{LM}}&LEGAL-BERT &4.47 &5.30 &4.85 &5.38 &7.42 &7.47 &10.9 &4.64 &8.61 &6.03 &6.03 &11.4 &11.3 &13.6\\
&MonoT5 &0.71 &0.65 &0.60 &0.79 &1.39 &1.41 &1.73 &0.38 &0.70 &0.49 &0.47 &1.17 &1.33 &0.61\\
&SAILER &16.6 &15.2 &14.0 &16.8 &17.2 &18.5 &25.1 &12.8 &23.7 &16.6 &17.0 &25.9 &25.3 &29.3\\
&PromptCase &17.1 &20.3 &18.5 &20.5 &35.1 &33.9 &38.7 &16.0 &29.7 &20.8 &21.5 &32.7 &32.0 &36.2\\
\midrule
\multirow{4}{*}{\rotatebox{90}{LLM}}&Mistral-7B &21.4 &25.4 &23.2 &25.7 &26.8 &28.5 &38.0 &16.0 &29.7 &20.8 &21.5 &21.9 &22.9 &30.8\\
&Qwen3-8B &21.6 &25.7 &23.5 &26.0 &26.7 &29.0 &37.8 &18.4 &34.1 &23.9 &25.2 &25.8 &27.5 &36.4\\
&Inf-7B &21.1 &25.1 &22.9 &25.5 &26.6 &28.7 &37.9 &19.0 &35.3 &24.7 &26.0 &26.5 &28.0 &37.6\\
\cmidrule{2-16}
\rowcolor{lightgray}
\cellcolor{white}&ReaKase-8B &\textbf{24.7} & \textbf{29.5} & \textbf{27.0} & \textbf{29.5} & \textbf{52.7} & \textbf{50.6} & \textbf{55.6} & \textbf{22.1} & \textbf{41.0} & \textbf{28.7} & \textbf{29.8} & \textbf{48.1} & \textbf{46.9} & \textbf{52.5} \\
\bottomrule
\end{tabular}}
\vspace{-0.3cm}
\end{table}

\subsection{Ablation Study (RQ2)}
To assess the contribution of legal entities and legal reasoning in \textsc{ReaKase-8B}, we perform an ablation study by selectively removing each component (Table~\ref{tab:ablation}). Removing both components leads to a substantial performance drop on COLIEE2022 and COLIEE2023, highlighting their critical roles in legal case retrieval. Incorporating legal entities alone yields modest gains over the baseline, whereas introducing legal reasoning alone provides a larger improvement, particularly in ranking metrics such as MRR@5 and NDCG@5. Combining both components achieves the best overall performance across most metrics, with Mi-F1 reaching 27.0 and 28.7, and NDCG@5 reaching 55.6 and 52.5 on COLIEE2022 and COLIEE2023, respectively. These results suggest that while legal reasoning contributes more substantially, the \textbf{integration of both the legal relation knowledge and the reasoning produces a complementary effect, leading to the most robust and consistent retrieval performance}.

\begin{table}[!h]\centering
\vspace{-0.8cm}
\caption{Ablation study. Entity and Reason denote legal entities and legal reasoning, separately. (\%)}\label{tab:ablation}
\resizebox{1\linewidth}{!}{
\begin{tabular}{c|c|ccccccc|ccccccc}
\toprule
\multirow{2}{*}{Entity} &\multirow{2}{*}{Reason} &\multicolumn{7}{c|}{COLIEE2022} &\multicolumn{7}{c}{COLIEE2023}\\
\cmidrule{3-16}
&  &P@5 &R@5 &Mi-F1 &Ma-F1 &MRR@5 &MAP &NDCG@5 &P@5 &R@5 &Mi-F1 &Ma-F1 &MRR@5 &MAP &NDCG@5 \\
\midrule
\xmark &\xmark &24.6 &29.2 &26.7 &29.0 &50.7 &48.5 &53.8 &20.7 &38.5 &27.0 &28.0 &44.9 &43.8 &48.6\\
\midrule
\cmark &\xmark &22.3 &26.4 &24.2 &26.5 &48.8 &46.5 &51.4 &21.7 &40.3 &28.2 &29.0 &46.8 &45.1 &50.7\\
\midrule
\xmark &\cmark &24.5 &29.1 &26.6 &29.1 &51.6 &49.1 &54.5 &22.0 &40.9 &28.6 &29.3 &47.1 &46.2 &52.0\\
\midrule
\cmark &\cmark &24.7 &29.5 &27.0 &29.5  &52.7 &50.6 &55.6 &22.1 &41.0 &28.7 &29.8 &48.1 &46.9 &52.5\\
\bottomrule
\end{tabular}}
\vspace{-1cm}
\end{table}

\subsection{Effectiveness of Reasoning Context (RQ3)}
To examine the impact of reasoning-context prompt templates on legal case retrieval, we evaluate three alternative user prompts within the contextualised legal case encoder on COLIEE2022 and COLIEE2023. This experiment investigates how prompt phrasing influences retrieval performance. In addition to the default template introduced in Section~\ref{sec:reakase}, two alternative prompts are illustrated in the below prompt figure, with corresponding results reported in Table~\ref{tab:effect_prompt}.

\begin{table}[!h]\centering
\vspace{-0.8cm}
\caption{Effectiveness of different user prompts. Prompts are shown below.}\label{tab:effect_prompt}
\resizebox{1\linewidth}{!}{
\begin{tabular}{c|ccccccc|ccccccc}
\toprule
\multirow{2}{*}{Templates} &\multicolumn{7}{c|}{COLIEE2022} &\multicolumn{7}{c}{COLIEE2023}\\
\cmidrule{2-15}
&P@5 &R@5 &Mi-F1 &Ma-F1 &MRR@5 &MAP &NDCG@5 &P@5 &R@5 &Mi-F1 &Ma-F1 &MRR@5 &MAP &NDCG@5 \\
\midrule
Prompt 1 &25.0 &29.8 &27.2 &29.9 &50.3 &48.3 &54.5 &21.6 &40.0 &28.0 &29.2 &48.8 &47.5 &52.8\\
Prompt 2 &24.5 &29.0 &26.6 &29.2 &49.7 &48.7 &54.2 &22.0 &40.8 &28.6 &29.8 &47.8 &46.7 &52.5\\
Default &24.7 &29.5 &27.0 &29.5  &52.7 &50.6 &55.6 &22.1 &41.0 &28.7 &29.8 &48.1 &46.9 &52.5\\
\bottomrule
\end{tabular}}
\vspace{-0.5cm}
\end{table}

Across both datasets, \textbf{all prompts demonstrate an effective contextualisation for case encoding}. On COLIEE2022, Prompt 1 achieves the strongest performance on classification metrics such as Mi-F1 and Ma-F1, while the default prompt performs best on ranking metrics including MRR@5, MAP, and NDCG@5. Prompt 2 consistently falls between the two. On COLIEE2023, performance differences narrow considerably, with most metrics differing by less than 0.5 points across prompts. Nevertheless, a similar trend emerges: Prompt 1 provides stable, balanced performance, while the default prompt achieves the best results on most metrics across both datasets. These findings suggest that concise label-style prompts (Default) sharpen ranking effectiveness and improve recall, whereas explicit instructional prompts (Prompt 1) deliver more consistent and generalizable gains across datasets.

\begin{center}
\begin{encode}[title={Prompt Template for Contextualised Case Encoding (Continued)}, label=prompt:recon]
\vspace{-1mm}
\hspace*{2em}\ \texttt{\# System Prompt}\\
\hspace*{2em}\ \texttt{The following contains key components of a legal case.} \\ \\
\hspace*{2em}\ \texttt{\# User Prompt 1}\\
\hspace*{2em}\ \texttt{Provide the key factual background:} $\{c_{\text{Fact}}\}$. \\ 
\hspace*{2em}\ \texttt{Provide the legal fact relation triplets:} $\{\mathcal{R}_{\text{Fact}}\}$. \\ 
\hspace*{2em}\ \texttt{Provide the key legal disputes:} $\{c_{\text{Issue}}\}$. \\ 
\hspace*{2em}\ \texttt{Provide the legal issue relation triplets:} $\{\mathcal{R}_{\text{Issue}}\}$. \\ 
\hspace*{2em}\ \texttt{Provide the legal reasoning between legal facts and legal}\\
\hspace*{2em}\ \texttt{issues:} $\{c_{\text{Reason}}\}$.\\ \\

\hspace*{2em}\ \texttt{\# Or User Prompt 2}\\
\hspace*{2em}\ \texttt{List the important legal facts as:} $\{c_{\text{Fact}}\}$. \\ 
\hspace*{2em}\ \texttt{List the important fact relations among events and parties}\\
\hspace*{2em}\ \texttt{as:} $\{\mathcal{R}_{\text{Fact}}\}$. \\ 
\hspace*{2em}\ \texttt{List the important legal issues as:} $\{c_{\text{Issue}}\}$. \\ 
\hspace*{2em}\ \texttt{List the important issue relations among events and parties}\\
\hspace*{2em}\ \texttt{as:} $\{\mathcal{R}_{\text{Issue}}\}$. \\ 
\hspace*{2em}\ \texttt{List the important legal reasoning as:} $\{c_{\text{Reason}}\}$.
\end{encode}
\end{center}

\subsection{Case Study}
To illustrate the important role of reasoning in \textsc{ReaKase-8B}, we compare a query case (090305) with two candidate precedents: a positive match (010294) and a negative match (005442), where 010294 is correctly retrieved only when reasoning is used, and 005442 is wrongfully retrieved without reasoning but correctly excluded when reasoning is incorporated.

As shown in the red text in below examples of generated legal reasoning, for query case 090305, which turns on whether coercion amounts to duress under Section 34(1)(f) of Canada’s Immigration and Refugee Protection Act (IRPA), a reasoning-aware model correctly retrieves case 010294. Both cases apply the same legal test from Ryan, whether the coercion meets the threshold of imminent peril, even though the outcomes differ. By contrast, a non-reasoning-based model retrieves 090305 with case 005442, which also cites Section 34(1) but focuses on procedural fairness and evidentiary flaws rather than duress. This shows that \textbf{reasoning alignment captures substantive relevance while filtering out statute-only matches}. This is because the statute-only matches are easily confused by the token level or wording level similarity that would otherwise lead to false positives.

\begin{center}
\begin{prompt}[title={Examples of Generated Legal Reasoning.}, label=prompt:recon]
\vspace{-1mm}
\hspace*{2em}\ \texttt{\# Generated Reasoning of Query Case (090305)}\\
\hspace*{2em}\ \texttt{The Officer used the correct legal tests: ``membership'' under s.34(1)(f) is broad; the complicity test (voluntary, significant, knowing contribution) applies to s.98/35(1)(a), not to membership; the SNM’s legitimacy is irrelevant. \textcolor{mydarkred}{\textbf{Duress requires imminent physical peril (Ryan). Mr. Mohamed showed general coercion, not imminent threats; thus his payments were not legally ``under duress.''}} On reasonableness, evidence supported that he was an SNM member. Therefore, inadmissibility was reasonably found; judicial review is dismissed. No certified question: no unsettled, dispositive legal issue.
} \\ \\
\hspace*{2em}\ \texttt{\# Generated Reasoning of Positive Match (010294)}\\
\hspace*{2em}\ \texttt{Applicant served in Iranian intelligence; officer found s.34(1) IRPA inadmissibility. Applicant claimed coercion/duress. Was the officer's decision reasonable, including treatment of duress? \textcolor{mydarkred}{\textbf{Membership finding under s.34(1)(f) was reasonably made. But the officer unreasonably failed to fully assess the duress claim (a material defense affecting culpable ``engaging/membership'')}}. Because the decision was unreasonable on duress, it was quashed and remitted to a different officer. Application granted.
}\\ \\
\hspace*{2em}\ \texttt{\# Generated Reasoning of Negative Match (005442)} \\ 
\hspace*{2em}\ \texttt{\textcolor{mydarkred}{\textbf{Apply reasonableness to the s.34(1) IRPA finding and no deference on procedural fairness}}. “Membership” requires more than suspicion: an institutional link/knowing participation, assessed contextually, considering exculpatory facts; low‑level, compelled tasks (pamphlets/soap in high school) alone are insufficient. The officer ignored relevant membership criteria and contrary evidence, relied on credibility concerns as proxy, and withheld influential CBSA/CTS advocacy memoranda, breaching fairness. Thus the decision was unreasonable and procedurally unfair, so it must be set aside and remitted to a different officer.}
\end{prompt}
\end{center}

\section{Conclusion}
This paper addresses two main challenges in legal case retrieval: limited legal reasoning and weak representation of structured legal relations. We introduce \textsc{ReaKase-8B}, which integrates legal element extraction and reasoning generation modules through contextualized prompting to enhance case semantics. \textsc{ReaKase-8B} achieves SOTA performance on both datasets, demonstrating its robustness and effectiveness for reasoning-oriented legal retrieval.

\section{Acknowledgements} This work is supported by Australian Research Council
CE200100025, DP230101196, DP230101753 and DE250100919.

\bibliographystyle{splncs04}
\bibliography{ref.bib}

\end{document}